\documentstyle[preprint,floats,epsf,eqsecnum,aps]{revtex}
\newcommand{\be}{\begin{equation}}
\newcommand{\ee}{\end{equation}}
\newcommand{\bea}{\begin{eqnarray}}
\newcommand{\eea}{\end{eqnarray}}

\def\del{\partial}
\begin{document}
\draft
\preprint{ \parbox{1.5in}{ \leftline{WM-99-109}
                           \leftline{JLAB-THY-99-16}                  
                         } 
         }
\title{Feynman-Schwinger representation approach to nonperturbative physics}
\author{\c{C}etin \c{S}avkl{\i}$^{1}$, John Tjon$^{2}$, Franz Gross$^{1,3}$  }
\address{ 
$^1$Department of Physics, College of William and Mary, Williamsburg, 
Virginia 23187\\
$^2$Institute for Theoretical Physics, University of Utrecht, Princetonplein 5,\\
P.O. Box 80.006, 3508 TA Utrecht, the Netherlands.\\
$^3$Jefferson Lab, 
12000 Jefferson Avenue, Newport News, VA 23606 }
\date{\today}
\maketitle
\begin{abstract}

The Feynman-Schwinger representation provides a convenient framework for the calculation of nonperturbative
propagators. In this paper we first investigate an analytically solvable case,
namely the scalar QED in 0+1 dimension. With this toy model we illustrate how 
the formalism works. The analytic result for the self energy is compared with 
the perturbative result. Next, using a $\chi^2\phi$  interaction, we discuss 
the regularization of various divergences encountered in this formalism. The 
ultraviolet divergence, which is common in standard perturbative field theory 
applications, is removed by using a Pauli-Villars regularization. We show that
the divergence associated with large values of Feynman-Schwinger parameter $s$
 is spurious and it can be avoided by using an imaginary Feynman parameter 
$is$.

\end{abstract}

\section{Introduction}
\label{introduction}

In nuclear physics one is often faced by problems that require nonperturbative
 methods. The best known example is the problem of bound states. Even if the underlying theory may have a small coupling constant (such as in QED), and therefore 
allows the use of perturbation theory in general, the treatment of bound states
are inherently nonperturbative. The n-body bound state is defined by the pole 
of the interacting n-body propagator. A perturbative approximation of 
n-body propagator does not produce the bound state pole location. This can most easily be 
seen by the following example:
\begin{equation}
\frac{1}{1-x}=1+x+x^2+x^3+\,.\,.\,.
\end{equation}
Any truncation of the righthandside (perturbation theory) will fail to produce
the pole which is located at $x=1$. Therefore it is essential that reliable
nonperturbative methods that take all orders of interaction into account are 
developed. For this reason, numerous nonperturbative methods have been developed and 
succesfully used in the literature. Some of the best known examples are 
relativistic bound state equations~\cite{NAKANISHI,TJON0,GROSS1}, and 
lattice gauge theory~\cite{ROTHE}. 

Relativistic bound state equations provide a practical and intuitive framework
to analyze bound states. However this practicality comes with certain 
drawbacks. In particular, the bound state equations in general lack gauge 
invariance. The second problem is associated with the fact that a completely 
self consistent solution of bound state equations is very difficult. A completely 
self consistent solution requires solving infinitely many coupled equations for all n-point functions of the theory. Since this is an impossible task, one 
is either forced to model various vertices and interaction kernels or 
specify them perturbatively.

The second and more recent  approach is known as lattice gauge theory (LGT). 
LGT is a Euclidean path integral based approach which 
relies on the discretization of space-time. An economical lattice simulation 
with a small lattice size of $5^4$ requires roughly $4\,({\rm links})\times 
8\,({\rm real\,\, SU(3)\,\, parameters})\times 5^4\,({\rm space-time points})=20000$ integrations. In general with larger lattice sizes this cost goes as 
$32N^4$. The disadvantage of discretization is twofold. The first one is 
the excessive computational time required for lattice simulations. The second
problem is the anomalies caused by the discretization, such as the fermion 
doubling problem~\cite{ROTHE}.   

In this paper we discuss yet another method known as the Feynman-Schwinger 
Representation (FSR) ~\cite{SIMONOV1,TJON1,BRAMBILLA}. Similar to the LGT, the FSR approach is also based on 
Euclidean path integral formalism. The basic idea in the FSR approach is to integrate out all fields at the expense of introducing quantum mechanical 
path integrals over the {\em trajectories of particles}. Replacing the path 
integrals over fields with path integrals over trajectories has an enormous
computational advantage. In the FSR approach, a calculation similar to the example given above requires only $4N$ integrations, where $N$ is now the steps 
a particle takes between the initial and final states. 
In addition to this enormous savings in computational time, the FSR approach 
also employs a space time continuum and therefore does not suffer from 
problems such as fermion doubling and the continuum limit. 

An additional motivation
in studying the FSR approach is to determine which subsets of diagrams give 
the dominant contribution to the n-body propagator. This is particularly
important in determining what kind of approximations are reasonable within 
the context of bound state equations. Therefore the FSR is a very promising 
tool to do this. 
In this paper we report on results for the 2-point function.
In studies of hadronic systems like $\pi N$ one usually models the
self-energy contribution through the lowest order one-loop graph. 
Also this is used as a starting point for the improved action
procedure proposed by Lepage~\cite{lepage}. It is clearly useful to
compare such a  lowest order approximation 
with the full results obtained from a FSR calculation.
We study here as a toy model the scalar QED (SQED) in 0 space and 1 time
dimension, which can be solved analytically.
The intriguing issue we also address in this paper is the difficulty found in
the Euclidean action functional for the case of a $\phi^3$-theory.
In applying the FSR to the 4-point function in the case of a
$\phi^3$-theory confined to generalized ladders one encounters a 
difficulty that the Euclidean action diverges. It
was conjectured and demonstrated in a simple example by Nieuwenhuis
\cite{TJON2} that this problem arises due to the continuation to
the Euclidean metric.
This problem is examined  in detail here and we in particular show 
that there exists a regularization method to remove this divergence. As a
result a clear prescription is given for dealing in a proper way with 
the Euclidean action in this case.
 
The organization of this paper is as follows: In the next section we 
start by discussing the case of SQED in 0+1 dimension. This is a simple 
case which can be worked out analytically. We consider the one-body 
and two-body propagators. We compare the one-body result for the dressed mass 
with the perturbation theory result. In the third section we consider the 
case of scalar interaction $\chi^2\phi$ in 3+1 dimension. We consider the 
issue of Wick rotation in Feynman parameter $s$, and present the FSR result 
for the one-body dressed mass obtained by Monte-Carlo integration. The 
result is again compared by the perturbation theory prediction.
   
\section{Scalar QED}
\begin{figure}
\begin{center}
\mbox{
   \epsfxsize=4.0in
   \epsfysize=1.5in 
\epsffile{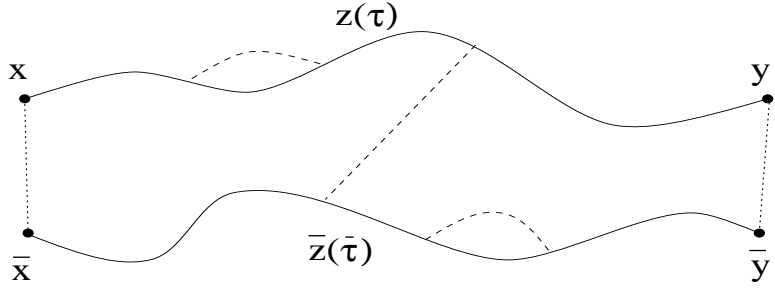}
}
\end{center}
\caption{A sample trajectory of each particle along with various interactions 
are shown.  }
\label{trajectory.plt}
\end{figure}

Massive scalar QED in 0+1 dimension is a simple interaction that enables one 
to obtain a fully analytical result for the dressed and bound state masses 
within the FSR approach. 
In this section we compare the self energy result obtained by 
perturbative methods with the full result obtained from the Feynman-Schwinger
 representation. The Minkowski metric expression for the scalar QED Lagrangian
in Stueckelberg form is given by
\begin{equation}
{\cal L}_{SQED}=-m^2\chi^2-\frac{1}{4}F^2+\frac{1}{2}\mu^2A^2-\lambda\frac{1}{2}(\partial A)^2+(\partial_\mu-ieA_\mu)\chi^*(\partial^\mu+ieA^\mu)\chi,
\end{equation}
where $A$ represents the gauge field of mass $\mu$, and $\chi$ is 
the charged field of mass $m$. We employ the Feynman gauge ($\lambda=1$). The 
presence of a mass term for the exchange field breaks the gauge invariance. 
Here the mass term was introduced in order to avoid infrared singularities 
which are present in 0+1 dimension. For dimensions larger than n=2 the 
infrared singularity does not exist and therefore the limit $\mu\rightarrow$ 
can be safely taken to restore the gauge invariance.    
Since we confine ourselves to 0+1 dimension, the antisymmetric tensor 
$F^{\mu\nu}$ vanishes. Therefore, in Euclidean metric and in 0+1 dimension, the
 scalar QED Lagrangian can be written as
\be
{\cal L}_{SQED}^E=\biggl[ m^2\chi^2+(\del\chi)^2 +\frac{1}{2}\,\mu^2 A^2+\frac{1}{2}\,(\del A)^2+ e^2\chi^2 A^2 - ieA(\chi^*\partial\chi-\chi\partial\chi^*)
\biggr].
\ee
In preparation for the path integration which will be performed below, it is 
more convenient to cast the Lagrangian into the following form
\begin{equation}
{\cal L}_{SQED}^E=\chi^*\biggl[m^2-\partial^2-2ieA\partial-ie\partial A+e^2A^2\biggr]\chi
+\frac{1}{2}A(\mu^2-\partial^2)A.
\end{equation}  
In order to construct a gauge invariant Green's function G, we introduce a 
gauge
link $U(x,y)$
\begin{equation}
U(x,y)\equiv {\rm exp} \left[-ie\int_x^ydz\,A(z)\right].
\end{equation} 
The two-body Green's function for the transition from the initial state
$\Phi_{i}=\chi^*(x)U(x,\bar{x})\chi(\bar{x})$ to final state 
$\Phi_{f}=\chi^*(y)U^*(y,\bar{y})\chi(\bar{y})$ is given by
\be
G(y,\bar{y}|x,\bar{x})=N\int {\cal D}\chi^*\int {\cal D}\chi\int {\cal D}A\,
\,\Phi^*_f\Phi_i\,e^{-S_E},
\label{g0.sqed}
\ee
where 
\be
S_E=\int d^4x \,\,{\cal L}_{SQED}^E.
\ee
Performing the path integrals over $\chi$ and $\chi^*$ fields one finds
\begin{eqnarray}
G(y,\bar{y}|x,\bar{x})&=&N\int {\cal D}A\,({\rm det}S)^{-1}U(x,\bar{x})U^*(y,\bar{y})\left[S(x,y)S(\bar{x},\bar{y})+S(x,\bar{x})S(y,\bar{y})\right]\nonumber\\
&&\,\,\,\,\,\,\,\,\,\,\,\,\,\,\,\,\,\,\,\,\times \,{\rm exp}\left[-\frac{1}{2}\int d^4z A(z)(\mu^2-\del^2)A(z)\right],
\end{eqnarray}
where the interacting propagator is defined by
\begin{equation}
S(x,y)\equiv <y\,|\,\frac{1}{m^2-\partial^2-2ieA\,\partial-ie\partial A+e^2A^2}\,|\,x>.
\end{equation} 
As in lattice gauge theory calculations we use the quenched approximation, ${\rm det}S\rightarrow 1$.   
In order to be able to carry out the remaining path integral over the exchange
field $A$ it is desirable to represent the interacting propagator in the 
form of an exponential. This can be achieved by using a Feynman representation for the interacting propagator. 
\begin{equation}
S(x,y)=<y\,|\int_0^{-i\infty}ds\,\, {\rm exp}\biggl[-s(m^2-\partial^2-2ieA\,\partial-ie\partial A+e^2A^2+i\epsilon)\biggr]|\,x>,
\label{phi3.int.prop}
\end{equation}
where the $s$ integration is along the imaginary axis. 
Let us now define
\begin{equation}
U(x,y,s)\equiv<y\,|\, {\rm exp}\biggl[-s(-\partial^2-2ieA\,\partial-ie\partial A+e^2A^2)\biggr]|\,x>
\end{equation}
where $U(x,y,s)$ satisfies
\begin{equation}
\frac{\partial}{\partial s}U(x,y,s)=(\del^2+2ieA\,\partial+ie\partial A-e^2A^2)U(x,y,s).
\end{equation}
This is equivalent to the Schroedinger equation for imaginary time $t=is$,
with Hamiltonian
\begin{equation}
H(p,z)=(p+ieA(z))^2.
\label{sqed.h}
\end{equation}
The matrix element of the interacting propagator Eq.~(\ref{phi3.int.prop}) can 
be written in terms of a quantum mechanical path integral. We know from quantum mechanics that
\begin{equation}
<y\,|\,{\rm exp}[-iH(q,p)\,t]\,|\,x>=\int {\cal D}q\,{\rm exp}\biggl[\,i\int_0^\infty {\cal L}(q(t),\dot{q}(t))\, dt\,\biggr].
\end{equation}
The Lagrangian for Eq.~(\ref{sqed.h}) is given by
\begin{equation}
L(z,\dot{z})=\frac{\dot{z}^2}{4}-ie\dot{z}A(z).
\end{equation}
Therefore, the quantum mechanical path integral representation for this propagator is given by
\begin{equation}
S(x,y)=-i\int_0^\infty ds\int {\cal D}z\, {\rm exp}\biggl[is(m^2+i\epsilon)-i/4\int_0^s d\tau\, \dot{z}^2(\tau)+ie\int_0^s d\tau\, \dot{z}A(z(\tau))\biggr],
\label{int.prop.sqed}
\end{equation}
where the boundary conditions are given by $z(0)=x$, $z(s)=y$. This 
representation allows one to perform the remaining path integral over the 
exchange field $A$. The final result for the two-body propagator 
involves a quantum mechanical path integral that sums up contributions coming 
from all possible {\em trajectories} of {\em particles}
\begin{equation}
G=-\int_0^\infty ds \int_0^\infty d\bar{s} \int ({\cal D}z)_{xy}\int ({\cal D}\bar{z})_{\bar{x}\bar{y}}\,e^{-S[Z]},
\label{g1.sqed}
\end{equation}
where $S[Z]$ is given by
\be
S[Z]\equiv -iK[z,s]-iK[\bar{z},\bar{s}]+V[z,\bar{z}].
\label{action}
\ee
The free and the interaction contributions to $S[Z]$ are given as
\bea
K[z,s]&=&(m^2+i\epsilon)s-\frac{1}{4s}\int_0^1 d\tau \,\frac{dz_\mu(\tau)}{d\tau}\frac{dz^\mu(\tau)}{d\tau},\label{k1}\\
V[z,\bar{z}]&=&+\frac{e^2}{2}\oint_Cd\tau\,\dot{z}(\tau)\oint_Cd\bar{\tau} \,\dot{\bar{z}}(\bar{\tau})\,\Delta(z(\tau)-\bar{z}(\bar{\tau}),\mu),
\label{v.sqed}\\
\Delta(x,\mu)&=& \int \frac{dp}{2\pi}\frac{e^{ip x}}{p^2+\mu^2}=\frac{e^{-\mu|z|}}{2\mu},
\label{kernel.sqed}
\eea
where $\Delta(x)$ is the interaction kernel. $K[z,s]$ represents the 
mass term and the kinetic term, and $V[z,s]$ 
includes the self energy and the exchange interaction 
contributions (shown in Fig.~\ref{trajectory.plt}). The contour of 
integration $C$ follows a counterclockwise trajectory $x\rightarrow y\rightarrow\bar{y}\rightarrow \bar{x}\rightarrow x$ as parameters $\tau$, and 
$\bar{\tau}$ are varied from 0 to $1$. The self energy and the exchange interaction contributions, which are embedded in expression~\ref{v.sqed}, have 
different signs. This follows from the fact that particles forming the two 
body bound state carry opposite charges.
The bound state spectrum can be determined from the spectral decomposition of the two body Green's function
\be
G(T)=\sum_{n=0}^{\infty}c_ne^{-m_nT},
\ee
where T is defined as the average time between the initial and final states
\be
T\equiv\frac{1}{2}(y_4+\bar{y}_4-x_4-\bar{x}_4).
\label{tdef}
\ee
In the limit of large $T$, the ground state mass is given by
\begin{equation}
m_0=\lim_{T\rightarrow\infty}-\frac{d}{dT} ln[G(T)]=\frac{\int {\cal D}Z S'[Z]e^{-S[Z]}}{\int {\cal D}Z e^{-S[Z]}},
\label{groundstate}
\end{equation}

\subsection{The one-body case}

In order to be able to compare with the perturbative result later, we 
concentrate on the one-body case. The one-body propagator is given by
\begin{equation}
G(0,T)=\int ds\int ({\cal D}z)_{0T}\, {\rm exp}\biggl[iK[z,s]-V_{0}[z]\biggr].
\end{equation}
The integral of the self interaction Eq.~(\ref{v.sqed}) can analytically be performed
\begin{eqnarray}
V_{0}[z]&=&\frac{e^2}{4\mu}\int_0^1 d\tau\, \dot{z}(\tau)\int_0^1 d\bar{\tau}\, \dot{z}(\bar{\tau})\, e^{-\mu|z(\tau)-z(\bar{\tau})|},\\
        &=&\frac{e^2T}{2\mu^2}\biggl[1-\frac{1-e^{-\mu T}}{\mu T}\biggr],
\end{eqnarray}
where the boundary conditions were chosen as $z(0)=0$, and $z(1)=T$.
Next, the path integral over $z$ can be evaluated after a discretization in proper time. Since the only path dependence in the propagator is in the kinetic term, the path integral over $z$ involves gaussian integrals which can be 
performed easily by using the following discretization
\be
({\cal D})_{0T}\rightarrow(N/4\pi s)^{N/2}\Pi^{N-1}_{i=1}\int dz_i.
\label{discreet}
\ee
The $s$ integral can also be evaluated by saddle point method giving
\begin{equation}
G(0,T)=N {\rm exp}\biggl[-mT-e^2\frac{T}{2\mu^2}+\frac{e^2}{2\mu^3}(1-e^{-\mu T})\biggr].
\end{equation}
This is an exact result for large times $T$. The dressed mass can easily be 
obtained by taking the logarithmic derivative of this expression. Therefore, 
the one-body dressed mass for SQED in 0+1 dimension according to the FSR 
formalism is given by
\begin{equation}
M=m+\frac{e^2}{2\mu^2}.
\label{sqed.fsr.mass}
\end{equation}
Simplicity of the SQED in 0+1 dimension also allows one to get an analytical result for the two-body bound state mass. It can easily be seen that the two-body 
result for the total mass is given by
\begin{equation}
M_{bound}=\left(m+\frac{e^2}{2\mu^2}\right)+\left(m+\frac{e^2}{2\mu^2}\right)-\frac{e^2}{\mu^2}=2m,
\label{sqed.fsr.bmass}
\end{equation}
where the first two terms are due to the one-body contributions and the
last term is due to the exchange interaction. The exchange contribution to 
the mass (up to a missing factor of two), was also reported earlier in 
Ref.~\cite{TJON2}. The interesting feature of the result in 
Eq.~(\ref{sqed.fsr.bmass}) is 
the fact that the positive shift of one-body masses are exactly compensated by
the negative binding energy created by the exchange interaction. Therefore the 
total bound state mass is exactly equal to the sum of {\em bare} masses.  
Therefore in this simple case vertex corrections do not contribute to the 
bound state mass.  

In order to be able to compare with the FSR prediction Eq.~(\ref{sqed.fsr.mass}) we consider the perturbative treatment of self energy.

\subsection{The perturbative result}
\begin{figure}
\begin{center}
\mbox{
   \epsfxsize=3.2in
\epsffile{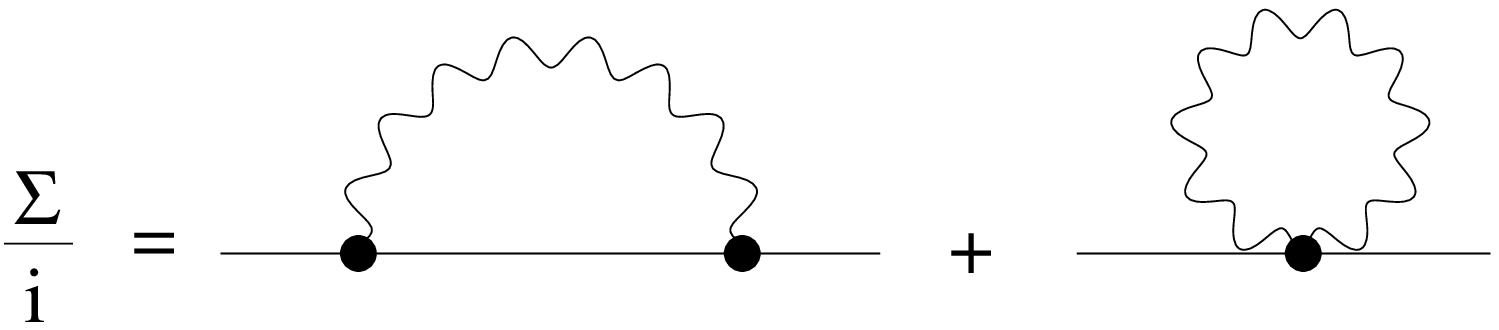}
}
\end{center}
\caption{ Self energy for the SQED.}
\label{sqed.self.fig}
\end{figure}

In this section we consider the perturbative treatment of the self energy
and  compare the perturbative result with the FSR prediction Eq.~(\ref{sqed.fsr.mass}). The self energy (Figure~\ref{sqed.self.fig}) in 0+1 dimension is given
 by 
\begin{equation}
\Sigma(p)=-ie^2\int_{-\infty}^{\infty}
\frac{dk}{2\pi}\frac{(2p-k)^2}{(k^2-\mu^2)[(p-k)^2-m^2]}+ie^2
\int_{\infty}^{\infty}\frac{dk}{2\pi}\frac{1}{k^2-\mu^2}.
\end{equation}
%
%----------------------------------------------
%----------------------------------------------
%
Performing the Wick rotation we get the following Euclidean expression
\begin{equation}
\Sigma(p)=-e^2\int_{-\infty}^{\infty}
\frac{dk}{2\pi}\biggl[\frac{(2p-k)^2}{(k^2+\mu^2)[(p-k)^2+m^2]}-\frac{1}{k^2+\mu^2}\biggr].
\end{equation}
Evaluating the integral we find
\begin{equation}
\Sigma(p)=-\frac{e^2}{2}\biggl[\frac{(i\mu-2p)^2}{\mu[m^2+(i\mu-p)^2]}+\frac{(im-p)^2}{m[\mu^2+(im+p)^2]}-\frac{1}{\mu}\biggr].
\end{equation}
The dressed propagator corresponding to this self energy is
\bea
-i\Delta_d(p)=&&\frac{-i}{m^2+p^2}+ \frac{-i}{m^2+p^2}\biggl\{ -i\Sigma_E(p)
\biggr\}\frac{-i}{m^2+p^2} + \cdots\nonumber\\
=&&\frac{-i}{m^2+p^2+\Sigma_E(p)}\, .
\eea
The coordinate space form of the dressed propagator is
\be
\Delta_d(t)=\int_{-\infty}^\infty
\frac{dp}{2\pi}\left(\frac{e^{ipt}}{m^2+p^2+\Sigma_E(p)}\right)
\simeq N e^{-Mt}\,  .
\ee
where $M$ is the dressed mass and $N$ is a normalization factor.
The dependence of
$M$ on the coupling strength $e$ can be obtained from the solution of the 
on-shell condition
\be
M=\sqrt{m^2+\Sigma_E(iM)},
\label{shellcondition}
\ee
which must be real if the dressed mass is to be stable.  
Therefore, for SQED, the equation determining the dressed mass takes the 
following form
\begin{equation}
M^2=m^2+\frac{e^2}{2}\biggl[\frac{(\mu-2M)^2}{\mu[m^2-(\mu-M)^2]}+\frac{(m-M)^2}{m[\mu^2-(m+M)^2]}+\frac{1}{\mu}\biggr].
\label{sqed.pert.mass}
\end{equation} 

\begin{figure}
\begin{center}
\mbox{
   \epsfxsize=4.8in
   \epsfysize=3.6in
\epsffile{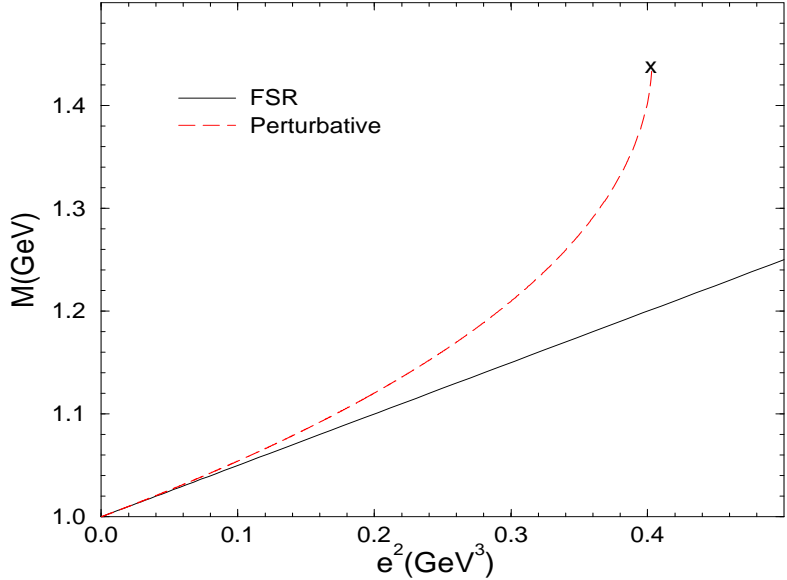}
}
\end{center}
\caption{The function $M(g^2)$ calculated by the FSR approach and the 
perturbative methods for values of $m=\mu=1$ GeV. According to the 
perturbative result there is a critical point at $g^2_{crit}=0.343$ (GeV)$^2$ beyond which the dressed mass 
becomes complex.}
\label{mvsg2.sqed}
\end{figure}

This perturbative result can be compared with the exact result Eq.~(\ref{sqed.fsr.mass}) found earlier. In figure~\ref{mvsg2.sqed} we present the case 
of $m=\mu=1$ GeV. For small values of coupling strength $e^2$ the perturbative 
and the full results converge. 
From the figure we see, that although the higher loop
contributions cannot entirely be neglected they are of limited size,
suggesting that in this case the lowest one-loop contribution
may be a reasonable approximation for not too strong couplings.
This is consistent with the results from Ref.~\cite{TJON2} in
the case of SQED in 2+1 dimension.
The perturbative result develops a complex mass
beyond a critical coupling $e^2_{crit}=0.343$ (GeV)$^2$. 
At the critical point a 'collision' takes place with another
real solution of  Eq.~(\ref{sqed.pert.mass}), leading to two
complex conjugated solutions with increasing $e^2$. This happens
at $M=1.49$ GeV.
This is an inadequacy of the perturbative approach. 

The occurrence of complex ghost poles in the propagator
has also been found in Lee-like models~\cite{kallen,ruijgrok} and in
$\pi-N$-interaction models~\cite{faassen}.
Moreover, it
is also interesting to note that a similar critical behaviour was 
also observed within the context of one body Dyson-Schwinger equation in 
Ref.~\cite{SAVKLI}. In the Dyson-Schwinger-Bethe-Salpeter studies of hadron 
structure the lack of real and finite mass poles in the quark propagator
is usually considered as an indication of confinement. On the other hand, the 
simple example of SQED study in 0+1 dimension shows that while the exact result for the dressed mass obtained from the FSR approach produces a {\em real} mass pole for all values of the coupling, the simple bubble sumation leads to {\em complex} mass poles for large coupling
values. Therefore the connection between confinement and lack of real mass 
poles is far from clear.     
%----------------------------------------------

%----------------------------------------------

Having presented the study of SQED in 0+1 dimension, where analytical results
are easily obtained and compared with the perturbative ones, we 
move on to the scalar $\chi^2\phi$ interaction in 3+1 dimension.

\section{The Feynman-Schwinger formalism for scalar fields}
\label{section3}

We consider the theory of charged scalar particles $\chi$ of mass $m$ interacting
through the exchange of a neutral scalar particle $\phi$ of mass $\mu$. 
For this case the analytical integration of path integrals are not possible
and one needs computational tools.  

The Euclidean Lagrangian for this theory is given by 
\be
{\cal L}_E=\chi^*\bigl[m^2-\del^2+g\phi\bigr]\chi+\frac{1}{2}\,\phi(\mu^2-\del^2)\phi.
\label{lagr0}
\ee
The two body Green's function for the transition from the initial state
$\Phi_{i}=\chi^*(x)\chi(\bar{x})$ to final state $\Phi_{f}=\chi^*(y)\chi(\bar{y})$ is given by
\be
G(y,\bar{y}|x,\bar{x})=N\int {\cal D}\chi^*\int {\cal D}\chi\int {\cal D}\phi
\,\Phi^*_f\Phi_i\,e^{-S_E}.
\label{g0}
\ee
Performing the path integrals over $\chi$ and $\chi^*$ fields one finds
\begin{equation}
G(y,\bar{y}|x,\bar{x})=N\int {\cal D}\phi\,({\rm det}S)^{-1}[S(x,y)S(\bar{x},\bar{y})+S(x,\bar{x})S(y,\bar{y})]e^{-\frac{1}{2}\int d^4z\phi(\mu^2-\del^2)\phi},
\end{equation}
where the interacting propagator is defined by
\begin{equation}
S(x,y)\equiv <y\,|\,\frac{1}{m^2-\del^2+g\phi}\,|\,x>.
\end{equation} 
In order to be able to carry out the remaining path integral over the exchange
field $\phi$ it is desirable to represent the interacting propagator in the 
form of an exponential. 
\begin{equation}
S(x,y)=<y\,|\int_0^{-i\infty}ds\, e^{-s(m^2-\del^2+g\phi+i\epsilon)}|\,x>.
\label{phi3.int.prop.2}
\end{equation}
Here we want to comment on a subtlety about the Feynman representation. In 
earlier works\cite{SIMONOV1,TJON1} the following Feynman representation was 
used 
\begin{equation}
S(x,y)\equiv<y\,|\int_0^\infty ds\, e^{-s(m^2-\del^2+g\phi)}|\,x>,\\
\end{equation}
The validity of this representation depends on the sign of the field $\phi$ which can be either positive or negative. If one accepts this representation, the
problem manifests itself as an exponentially diverging $s$ dependence after the path integral over $\phi$ is performed. In order to circumvent the problem of the $s$ singularity we use the Feynman representation given in Eq.~(\ref{phi3.int.prop.2}). Let us define
\begin{equation}
U(x,y,s)\equiv<y\,| e^{-s(-\del^2+g\phi)}|\,x>,
\end{equation}
where $U(x,y,s)$ satisfies
\begin{equation}
\frac{\partial}{\partial s}U(x,y,s)=(\del^2-g\phi)U(x,y,s).
\end{equation}
This is equivalent to the Schroedinger equation for imaginary time $t=is$,
with Hamiltonian
\begin{equation}
H(p,z)=p^2-g\phi(z).
\label{phi3.h}
\end{equation}
The Lagrangian for the Hamiltonian given in Eq.~(\ref{phi3.h}) is
\begin{equation}
L(z,\dot{z})=\frac{\dot{z}^2}{4}+g\phi(z).
\label{phi3.l}
\end{equation}
Therefore, the interacting propagator can be expressed as
\begin{equation}
S(x,y)=-i\int_0^\infty ds\int {\cal D}z\, {\rm exp}\biggl[is(m^2+i\epsilon)-i/4\int_0^s d\tau\, \dot{z}^2(\tau)+ig\int_0^s d\tau\, \phi(z(\tau))\biggr],
\end{equation}
where the boundary conditions are given by $z(0)=x$, $z(s)=y$.
The final result for the two-body propagator 
involves a quantum mechanical path integral that sums up contributions coming 
from all possible {\em trajectories} of {\em particles}.
The only difference from the SQED case Eq.~(\ref{int.prop.sqed}) is the 
replacement of $e\dot{z} A(z(\tau)) \rightarrow g\phi$. Therefore,
for the two body propagator one arrives at the same expresion as Eq.~(\ref{g1.sqed}) except 
now the new definition (compare with Eq.~(\ref{v.sqed})) of
the interaction term is 
\bea
V[z,\bar{z},s,\bar{s}]&=&V_{0}[z,s]+2V_{12}[z,\bar{z},s,\bar{s}]+V_{0}[\bar{z},\bar{s}],\label{vphi3}
\eea
where
\bea
V_{0}[z,s]&=&\frac{g^2}{2}s^2\int_0^1d\tau\int_0^1d\tau' \,\Delta(z(\tau)-z(\tau'),\mu),
\label{v0.phi3}\\
V_{12}[z,\bar{z},s,\bar{s}]&=&\frac{g^2}{2}s\bar{s}\int_0^1d\tau\int_0^1d\bar{\tau}\, \Delta(z(\tau)-\bar{z}(\bar{\tau}),\mu).\label{v12.phi3}
\eea
Here the $V_{0}[z,s]$ term represents the self energy contribution, while
the $V_{12}[z,\bar{z},s,\bar{s}]$ term represents the exchange interaction (Fig.~\ref{trajectory.plt}).
The notable difference compared to the SQED case is that the interaction terms now depend on the $s$ variable. The second difference from the SQED case is the fact that self energy and exchange interaction terms have the same signs. 
The interaction kernel $\Delta(x)$ is defined by
\bea
\Delta(x,\mu)&=& \int \frac{d^4p}{(2\pi)^4}\frac{e^{ip\cdot x}}{p^2+\mu^2}=\frac{\mu}{4\pi^2|x|}K_1(\mu|x|).
\label{kernel.phi3}
\eea
The time of propagation, T, is defined as before in Eq.~(\ref{tdef})

In principle one can work with equation~(\ref{groundstate}), using the 
interaction given in Eq.~(\ref{vphi3}), to determine the ground state energy of 
the bound state. However this is in practice very costly. The problem is the 
oscillatory behavior of the integrand which forbids the use of Monte-Carlo 
techniques for integration. Therefore it is desirable to make a Wick rotation 
in variable $s$. In the next section we discuss how this rotation can be made 
without leading to a large s divergence associated with the interaction term.
    
\subsection{The large s behavior and Wick rotation}

\begin{figure}
\begin{center}
\mbox{
   \epsfxsize=4.8in
\epsffile{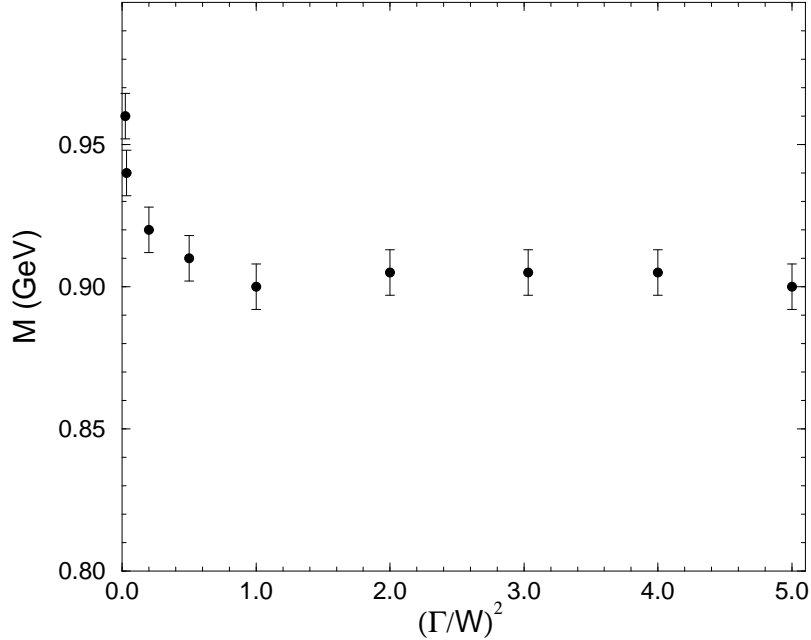}
}
\end{center}
\caption{The figure shows the insensitivity of the dressed mass to the 
width $\Gamma$ for $\Gamma\ge W$. This case represents $g=5$ GeV, and $mT=40$.
Results were obtained averaging about 4 Monte-Carlo runs at each $\Gamma$.}
\label{gamma}
\end{figure}

The one body propagator is given by 
\be
G=i \int_0^\infty ds\,\int {\cal D}z\,\, {\rm exp}\left[im^2s-i\frac{k^2}{4s}-s^2v - \epsilon s\right],
\label{spart}
\ee
where the s-independent functionals $k^2[z]$ and $v[z]$ are defined by
\bea
k^2[z]&\equiv& \int_0^1 d\tau \,\frac{dz_\mu(\tau)}{d\tau}\frac{dz^\mu(\tau)}{d\tau},\label{k2}\\
v[z]& \equiv & \frac{g^2}{2}\int_0^1d\tau\int_0^1d\tau' \,\Delta(z(\tau)-z(\tau'),\mu).\label{v}
\eea
The path integral is discretized using
\be
({\cal D})_{xy}\rightarrow(N/4\pi s)^{2N}\Pi^{N-1}_{i=1}\int d^4z_i,
\ee
where the s-dependence is critical in obtaining the correct normalization.
The one body propagator after this discretization is given by 
\begin{equation}
G=i\left(\frac{N}{4\pi}\right)^{2N}\int \Pi_{i=1}^{N-1}dz_i \int_0^\infty \frac{ds}{s^{2N}}\,{\rm exp}\biggl[im^2s-i\frac{k^2}{4s}-s^2v\biggr],
\label{g1.phi3}
\end{equation}
This is a well defined and finite integral. In principle the number of steps
$N$ should be taken to infinity. If one keeps the N fixed, a simple 
replacement of $s\rightarrow i\,s$ clearly leads to a divergent $s$ integral and is therefore 
not allowed. In order to put this integral into a form that allows Wick 
rotation without changing the physics we use the following trick. At large 
values of s the integral in Eq.~(\ref{g1.phi3}) is highly damped because of 
the $v$ and $s^{2N}$ terms. The integrand is also highly oscillatory as $s\rightarrow 0$, or $s\rightarrow \infty$  and therefore these regions do not 
contribute to the integral. In the limit $g^2\rightarrow 0$ the dominant 
contribution to the integral in Eq.~(\ref{g1.phi3}) can be shown, by using the
saddle point method, to come from
\be
s=is_0=i\frac{k}{2m}.  
\ee
Since the large $s$ values do not contribute to the integral even without
 the interaction term, it is a good approximation to suppress the $g^2$ term 
at large s values. While this suppression is done it is important that 
the integrand is not modified in the region of dominant contribution 
$s\sim is_0$. This can be achieved by scaling the $s$ variable, in the interaction term only, by 
\begin{equation}
s\rightarrow \frac{s}{R(s,s_0)},
\label{sredef}
\end{equation} 
where
\begin{equation}
R(s,s_0)\equiv 1-(s-is_0)^2/\Gamma^2.
\end{equation}
In the free case, $(g^2=0)$, the width $W$ of the region of dominant $s$ 
contribution goes as 
\be
W=\sqrt{\frac{T}{2m^{3}}}.
\ee
Therefore, in the free case the dominant contribution to the $s$ integral 
comes from $i(s_0-W)<s<i(s_0+W)$. In order to ensure that the scaling given in 
Eq.~(\ref{sredef}) does not make a significant change in the region of 
dominant contribution, $\Gamma$ should be chosen such that
\be
\Gamma\ge W
\ee
It should be pointed out that as one increases the coupling strength, the value
of $s_0$ will deviate from its free value. Therefore, in general, $s_0$ has to
be defined self consistently by monitoring the peak of the $s$ distribution.
In Figure~(\ref{gamma}) we display the insensitivity of the dressed mass to 
the width $\Gamma$, for $g=5$ GeV and $mT=40$. The results we present in the
remainder of the paper are obtained with a choice of $\Gamma^2=2W^2$.
%
%------------------------------------------------------------------- 
%---------------------------------------------------------------------
%
As a result of the scaling Eq.~(\ref{sredef}) the interaction term disappears 
at large $s$ values where the integrand does not contribute anyway. The 
benefit of this replacement is in the fact that even though N is kept 
finite one can perform a Wick rotation rigorously in variable $s$ to find
a nonoscillatory and {\em finite} integral. Now let us take a closer look at the Wick rotation.

After the redefinition given in Eq.~(\ref{sredef}) the $s$ integral in 
Eq.~(\ref{spart}) takes the following form 
\be
G\propto \int_0^\infty \frac{ds}{s^{2N}}\,\, {\rm exp}\biggl[im^2s-i\frac{k^2}{4s}-g^2 s^2/R^2(s)\,v\biggr].
\label{spartnew}
\ee
The exponent has a singularity in the complex $s$ plane at $s_p=is_0\pm\Gamma$. One of these singularities is on the path of the Wick rotation. However it can easily be seen that it does 
not contribute to the integral. In particular, the contribution of the 
singularity at $s_s=is_0+\Gamma$, let's call it $G_{pole}$, is given by
\begin{equation}
G_{pole}=i\,\frac{e^{im^2 s_p-ik^2/(4s_p)}}{s_p^{2N}}\lim_{\delta\rightarrow 0}\delta\int_0^{2\pi} d\theta\, {\rm exp}\biggl[i\theta-g^2s_p^2\Gamma^2/(4\delta^2) e^{-2i\theta}\,v\biggr],
\end{equation}  
which is identically equal to 0.
The contribution to the integral coming from the quarter circle can also be 
shown to vanish. On the quarter circle the interaction term approaches zero,
and the integral is dominated by $1/s^{2N}$ term which vanishes as the radius
of circle goes to infinity. Therefore the integral vanishes along the quarter 
circle. The vanishing of the interaction term on the quarter circle is only 
possible if one assumes that the radius of the quarter circle is much greater 
than $\Gamma$. Therefore $\Gamma$ can not be taken to infinity until the s 
integral is performed. 
%
%------------------------------------------------------------------- 
\begin{figure}
\begin{center}
\mbox{
   \epsfxsize=3.0in
\epsffile{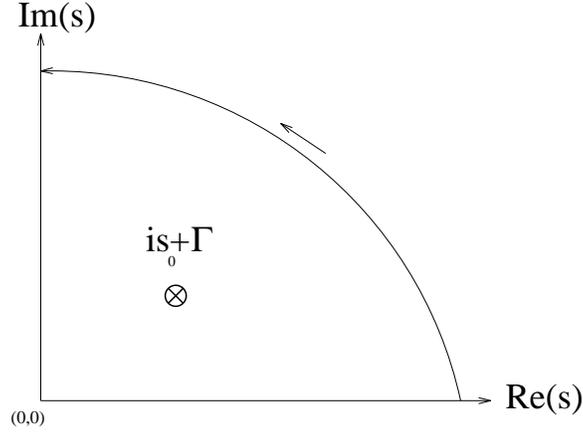}
}
\end{center}
\caption{Wick rotation}
\label{wick}
\end{figure}
%---------------------------------------------------------------------
%

Thus, one can indeed perform the Wick rotation without any complication
to find a finite and nonoscillatory expression for the fully interacting two-body propagator:  

\begin{eqnarray}
G&=&\int_0^{\infty} ds \int_0^{\infty} d\bar{s} \int ({\cal D}z)_{xy}\int ({\cal D}\bar{z})_{\bar{x}\bar{y}}\nonumber\\
&&\times{\rm exp}\biggl[-K[z,s]-K[\bar{z},\bar{s}]+V_0[z,s_r]+V_0[\bar{z},\bar{s}_r]+2V_{12}[z,\bar{z},s_r,\bar{s}_r]\biggr],
\label{g2}
\end{eqnarray}
where
\be
s_r\equiv \frac{s}{R(s,s_0)}.
\ee
The discretized versions of kinetic and interaction terms are given by
\bea
K[z,s]&\rightarrow& (m^2+i\epsilon)s-\frac{N}{4s}\sum_{i=1}^{N}(z_i-z_{i-1})^2,\\
V_0[z,s]&\rightarrow&\frac{g^2s^2}{2N^2}\sum_{i,j=1}^{N}\Delta(\frac{1}{2}(z_i+z_{i-1}-z_j-z_{j-1}),\mu),\\
V_{12}[z,\bar{z},s,\bar{s}]&\rightarrow&\frac{g^2s\bar{s}}{2N^2}\sum_{i,j=1}^{N}\Delta(\frac{1}{2}(z_i+z_{i-1}-\bar{z}_j-\bar{z}_{j-1}),\mu).
\eea
Having outlined the treatment of the large s behavior and the Wick rotation, 
we next address the regularization of the ultraviolet (short distance) 
singularities.

\subsection{The ultraviolet regularization}

The ultraviolet singularity in the kernel $\Delta(x,\mu)$ Eq.~(\ref{kernel.phi3}) 
can be regularized using a Pauli-Villars regularization prescription. In order
to do this one replaces the kernel
\be
\Delta(x,\mu)\longrightarrow \Delta(x,\mu)-\Delta(x,\alpha\mu),
\ee
where $\alpha$ is in principle a large constant. The ultraviolet
singularity in the interaction is of the type
\be
\int dz\, z\, \Delta(z,\mu).
\ee
At short distances the kernel $\Delta(z,\mu)$ goes as $1/z^2$. Therefore,
we have a logarithmic type singularity. The Pauli-Villars regularization
takes care of this singularity. The Pauli-Villars regularization 
is particularly convenient for Monte-Carlo simulations since it only 
involves a modification of the kernel. In order to achieve an efficient
convergence in numerical simulations we use a rather small cut-off parameter 
$\alpha=3$. This choice leads to a less singular kernel. However this is not a
major defect since the value of $\alpha$ can be increased arbitrarily at the 
cost additional computational time. 

\subsection{Perturbation theory result for self energy}

In this section we study the self energy of a particle of mass $m$ in lowest
order in perturbation theory.  We carry out the study in 3+1
dimension.

\begin{figure}
\begin{center}
\mbox{
   \epsfxsize=2.5in
   \epsfysize=0.7in
\epsffile{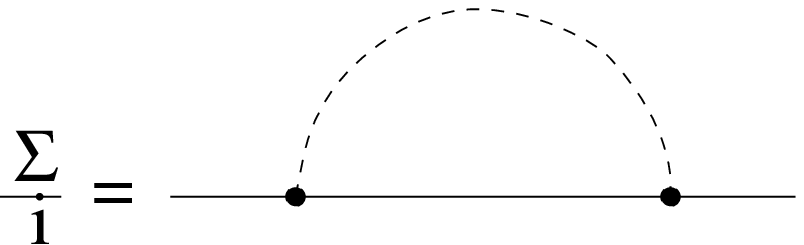}
}
\end{center}
\caption{The self energy for $\chi^2\phi$ interaction.}
\label{phi3.self.fig}
\end{figure}

The lowest order ``bubble'' diagram is shown in Fig.~\ref{phi3.self.fig}.  In 3+1
dimensions this diagram is
\be
\Sigma(p)=ig^2\int
\frac{d^4k}{(2\pi)^4}\frac{1}{\left(m^2-(p-k)^2\right)\left(\mu^2-k^2\right)}
\, . \label{eqn1.1}
\ee
%
%----------------------------------------------

%----------------------------------------------
%
In order to compare the perturbative result with the FSR prediction we use the same regularization method, namely the Pauli-Villars regularization.
Assuming that $p^2<(m+\mu)^2$, the integration over $k$ may be
rotated from the real axis to the imaginary axis without meeting any
singularities.  Substituting $k\to-ik$ and $p\to -ip$ and
gives the Euclidean expression for the self energy
\begin{equation}
\Sigma_E(p)=-g^2\int
\frac{d^4k}{(2\pi)^4}\frac{\Lambda^2-\mu^2}{\left(m^2+(p-k)^2\right)\left(\mu^2+k^2\right)\left(\Lambda^2+k^2\right)},
\end{equation}
where the Pauli-Villars regularization mass $\Lambda$ is chosen to be $\Lambda=3\mu$.
Using the Feynman trick the integral can be evaluated giving
\be
\Sigma_E(p)=-\frac{g^2}{16\pi^2}\left[I(p^2,\mu)-I(p^2,\Lambda)\right],
\label{sigphi3}
\ee
where $I(p^2,\mu)$ is defined by
\begin{eqnarray}
I(p^2,\mu)&\equiv&\int_0^1 dx \,{\rm ln}[m^2 x+p^2 x(1-x)+\mu^2(1-x)],\nonumber\\
&=&-\frac{1}{2}\biggl[\,2\, D \,{\rm tan}^{-1}\left(\frac{\mu^2-m^2-p^2}{D}\right)+(\mu^2-m^2-p^2)\,{\rm ln}(\frac{\mu^2}{m^2})\nonumber\\
&&\,\,\,\,\,\,\,\,-2\,D\,{\rm tan}^{-1}\left(\frac{\mu^2-m^2+p^2}{D}\right)-2p^2[-2+\,{\rm ln}(m^2)]\biggr],
\end{eqnarray}
where
\begin{equation}
D\equiv\sqrt{-\mu^4+2\mu^2(m^2-p^2)-(m^2+p^2)^2}.
\end{equation}
The dependence of
$M$ on the coupling strength $g$ can be obtained by 
analytic continuation of the Euclidean form 
of $\Sigma$ given in Eq.~(\ref{sigphi3}).  
It is found from the on-shell condition Eq.~(\ref{shellcondition}) that
\be
M^2=m^2-\frac{g^2}{16\pi^2}\left[I(-M^2,\mu)-I(-M^2,\Lambda)\right],
\label{phi3.pert.mass}
\ee
The mass
$M$ is therefore the solution of this equation
which must be real if the dressed mass is to be stable.  

\begin{figure}
\begin{center}
\mbox{
   \epsfxsize=4.8in
\epsffile{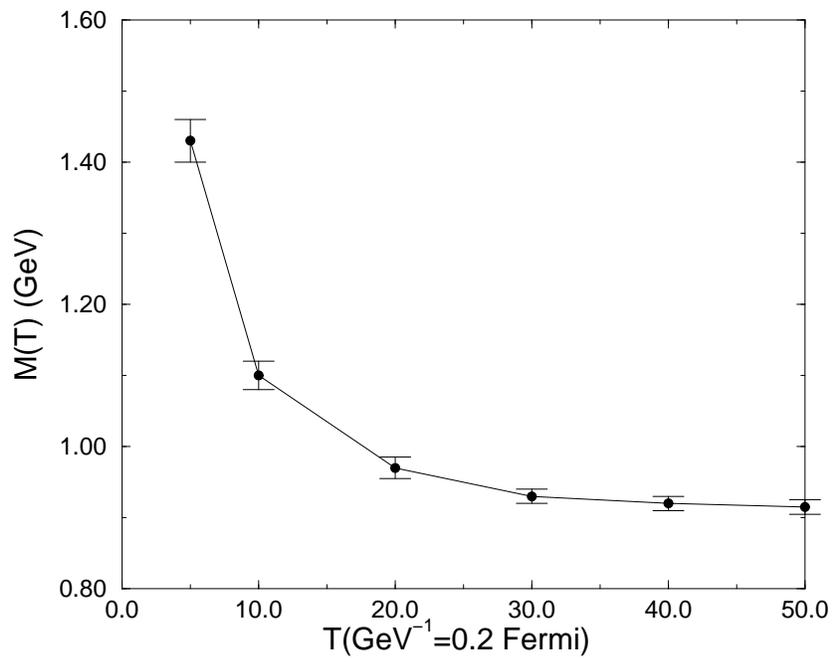} }
\end{center}
\caption{The Monte-Carlo result for the function $M(T)$ 
 is shown. Error bars reflect statistical errors associated with the 
Monte-Carlo sampling. The plateau region occurs around $T\ge 40$.}
\label{mvst.phi3}
\end{figure}

The FSR result is obtained through Monte-Carlo integration. The dressed 
mass $m_0\equiv \lim_{T\rightarrow\infty}M(T)$, which is given by 
Eq.~(\ref{groundstate}) becomes largely 
independent of $T$ at large times $mT\ge 40$. As the coupling strength is 
increased the plateau is shifted towards higher $T$ values. In 
Figure~\ref{mvst.phi3} we demonstrate how the stability is achieved as $T$ 
increases for the case of $g=5$ GeV.
%
%----------------------------------------------

%----------------------------------------------
%
%
%----------------------------------------------
\begin{figure}
\begin{center}
\mbox{
   \epsfxsize=4.8in
   \epsfysize=3.6in
\epsffile{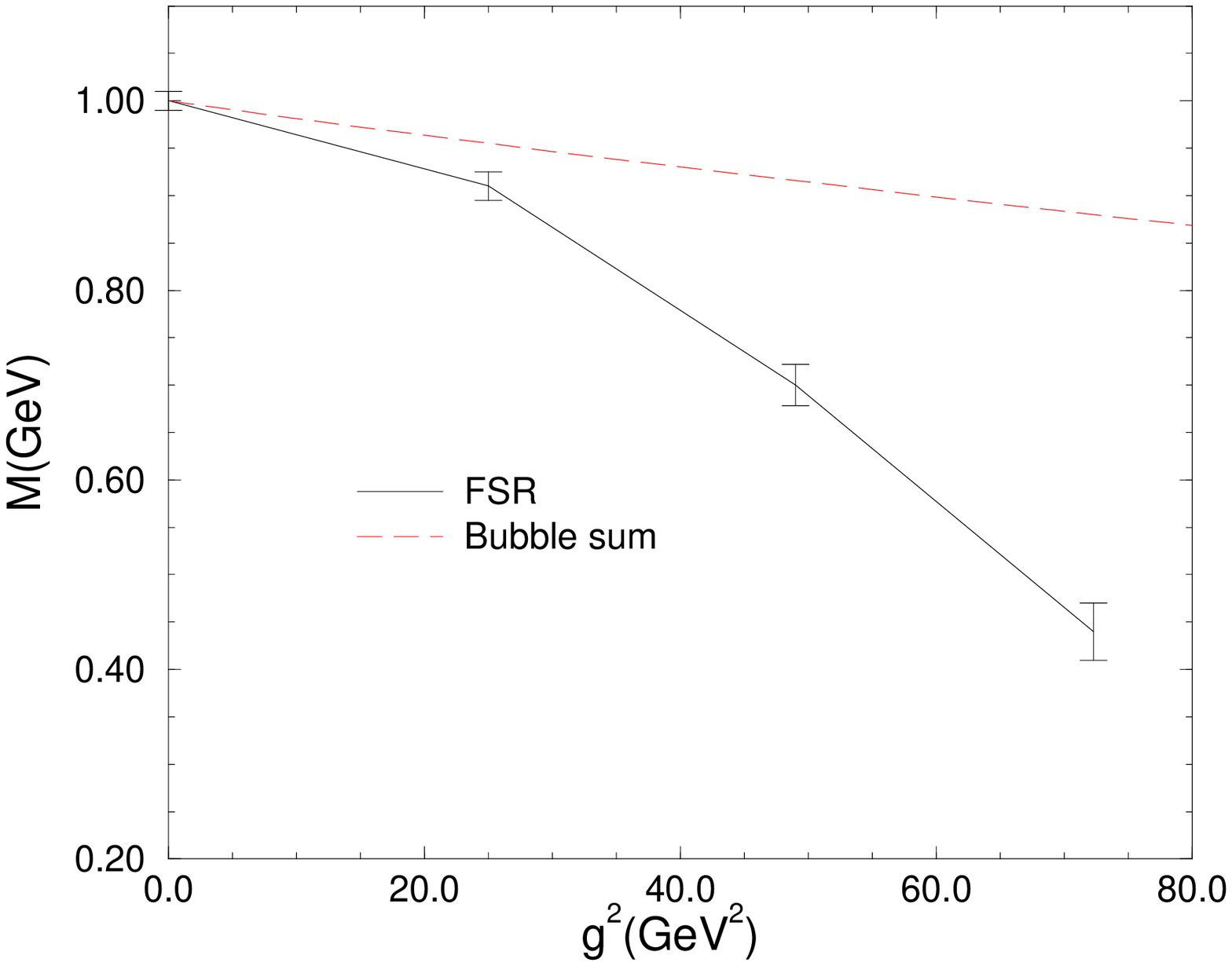}
}
\end{center}
\caption{The FSR result for the function $M(g^2)$ obtained from the 
Monte-Carlo simulation is shown along with the perturbative result. Error
bars on the Monte-Carlo result are due to the fluctuations of the correlation function (which goes as $e^{-mt}$) under time. }
\label{mvsg2.phi3}
\end{figure}
%----------------------------------------------
%
The behavior of dressed mass $M(g^2)$ as a function of the coupling constant 
is illustrated in Fig.~\ref{mvsg2.phi3}.  $M(g^2)$ is always smaller than unity, 
and decreases 
as $g$ increases. The agreement of the FSR result with the perturbative 
prediction is very good at low $g^2\le 30$ (GeV)$^2$. 
We see that the mass shift is negative, corresponding to an attractive
interaction. This should be contrasted with the SQED,
where a positive mass shift is predicted.
From the figures we see that the higher loop contributions
increase the mass shifts in both cases.

\begin{figure}
\begin{center}
\mbox{
   \epsfxsize=4.8in
   \epsfysize=3.6in
\epsffile{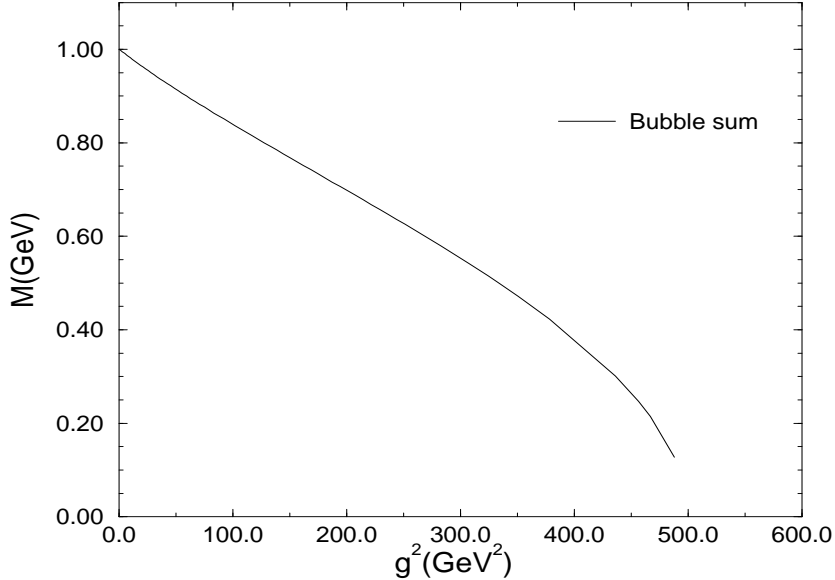}
}
\end{center}
\caption{The perturbative result for the function $M(g^2)$ is shown for $m=1$ GeV, and $\mu=0.15$. The critical 
coupling is given by $g_{\rm crit}=22.2$ GeV. For larger values of $g$ there are no real solutions}
\label{mvsg2.phi3.crit}
\end{figure}

Moreover, the perturbative result displays a critical point as in 
the 0+1 dimension SQED case.  
According to the perturbative 
result (see Fig.~\ref{mvsg2.phi3.crit}), the dressed mass $M$ decreases 
up to a critical critical value $g_{\rm crit}$ which occurs when 
the mass reaches to $M_{\rm crit}=0.094$ GeV. For this simple case 
the critical coupling is given by $g_{\rm crit}=22.2$ GeV. For larger 
values of $g$ there are no real solutions, 
showing that the dressed particle is unstable.  {\it For $g>g_{\rm crit}$ the 
state does not propagate as a free particle.\/}  For the example shown in the 
figure, $m=1$ GeV, and $\mu=0.15$ GeV.  
%
%----------------------------------------------
%----------------------------------------------
%

\section{conclusions}
In this paper we have considered the SQED interaction in 0+1 dimension and the scalar 
$\chi^2\phi$ interaction in 3+1 dimension. We have shown that for the SQED, the 
analytical FSR result and the perturbative one are in agreement at small 
couplings. The exact SQED result for the dressed mass is real while a lowest
order bubble sum produces complex mass poles as the coupling constant is increased.
This example shows that the lack of real and finite mass poles do not 
necessarily imply confinement unless they are obtained by fully 
nonperturbative calculations. 

For the $\chi^2\phi$ interaction we have shown that it is possible to perform 
a Wick rotation in Feynman parameter s and obtain a convergent expression for 
the s integration. 

\section{acknowledgement}
This work was supported in part by the US Department
of Energy under grant No.~DE-FG02-97ER41032.
One of us (JT) likes to thank the theory group of the Jefferson
Lab for the hospitality extended to him, where this work was
performed.

\end{document}